# First principles calculations of oxygen adsorption on the UN (001) surface

Yu.F. Zhukovskii[1], D. Bocharov[1*], E.A. Kotomin[2], R.A. Evarestov[3] and A.V. Bandura[3]

[1]*Institute for Solid State Physics, Kengaraga 8, LV- 1063 Riga, Latvia*
[2]*European Commission, Joint Research Centre, Institute for Transuranium Elements, Hermann von Helmholtz Pl. 1, D-76344 Eggenstein-Leopoldshafen, Germany*
[3]*Department of Quantum Chemistry, St. Petersburg State University, 198504 St. Petergof, Universitetsky Prosp. 26, Russian Federation*

*PACS numbers:* 68.43.Bc, 73.20.At, 71.15.Mb

Fabrication, handling and disposal of nuclear fuel materials require comprehensive knowledge of their surface morphology and reactivity. Due to unavoidable contact with air components (even at low partial pressures), UN samples contain considerable amount of oxygen impurities affecting fuel properties. The basic properties of O atoms adsorbed on the UN(001) surface are simulated here combining the two first principles calculation methods based on the plane wave basis set and that of the localized atomic orbitals.

The actinide nitrides and carbides, *e.g.,* uranium mononitride (UN) with a face centered cubic (*fcc*) rock salt structure, belong to the family of non-oxide ceramic nuclear fuels considered as promising candidates for the use in Generation-IV fast nuclear reactors. These materials reveal several advantages over traditional $UO_2$ fuel (*e.g.*, higher thermal conductivity and metal density) [1]. One of the problems with nitride and carbide fuels is their active interaction with the oxygen which results in an effective fuel oxidation and degradation [2]. This could affect the fabrication process as well as the fuel performance and safety. First experimental studies on O in UN were performed in 80ies ([1] and references therein). These activities were continued recently combining several techniques ([2] and references therein). However, understanding of the atomistic mechanism of fuel oxidation needs first principles theoretical modeling. Thus, to shed more light on this problem, we study here theoretically the interaction of atomic oxygen with the UN(001) surface.

Theoretical simulations of uranium compounds are especially complicated due to a relativistic character of an electron motion in the U atomic core and the strong electron-electron correlation. Moreover, UN is characterized by a mixed metal-covalent chemical bonding. Physical and chemical properties of light actinides are determined by partly localized *5f* electrons, which determine a number of properties, such as mixed valence, magnetism, *etc.* A series of first principles DFT calculations on pure and defective $UO_2$ were performed recently (*e.g.*, [3-8]) whereas a number of similar calculations on the nitride fuels is still much more limited [9-15]. In our recent paper [15] the methodology was proposed for LCAO calculations of the UN surface properties. The first results on the pure UN surfaces were presented therein using two approaches based on the basis sets of atomic orbitals (AO) and plane waves (PW), respectively. Use of the two different methods greatly increases the reliability of the results obtained.

To simplify modeling of the oxygen interaction with UN powder surface, we study here only the (001) surface which according to Tasker [16] has the lowest energy. To simulate the perfect UN(001) substrate as well as its interaction with oxygen, we have employed the DFT-PW computer code *VASP 4.6* [17] based on the use of a plane wave basis set and the method of projector-augmented-waves (PAW) for atomic core description. We apply the non-local exchange-correlation functional Perdew-Wang-91 using the generalized gradient approximation (GGA) [18] and the scalar relativistic PAW pseudopotentials representing the U core electrons (with $6s^26p^66d^25f^27s^2$ valence shell), N ($2s^22p^3$) and O ($2s^22p^4$) atoms (containing 14, 5 and 6 valence electrons, respectively). The cut-off energy has been chosen to be 520 eV. We use the Monkhorst-Pack scheme [19] with mainly 8×8×1 *k*-point meshes in the Brillouin zone (BZ).

---
[*]Corresponding author: e-mail: bocharov@latnet.lv

As the second method, we have used the *CRYSTAL-06* computer code [20] based on the Gaussian-type functions centered on the atomic nuclei as the basis sets for expansion of the linear combination of atomic orbitals (LCAO). We use the non-local exchange-correlation functional PBE [21]. The oxygen basis set (BS) 8-411G(1*d*) was taken from Ref. [22]. For the N atom, the all-electron BS 6-311G(2*d*) has been used [23]. Finally, for the U atom we have used the energy-adjusted relativistic small core (60 electrons in core) effective potential from Ref. [24]. To get rid of the basis set linear dependence in the *CRYSTAL* LCAO calculations, the diffuse *s*-, *p*-, *d*- and *f*- Gaussian-type orbitals with exponents < 0.2 a.u.$^{-1}$ have been removed from the basis sets. The exponents of other polarization functions have been reoptimized, to restore the required precision in the total energy. High accuracy in both *k*-set mesh and DFT integration grid (XLGRID) has been applied for all *CRYSTAL-06* calculations. Prior to a study of surface properties, the bulk structure optimization of UN crystal has been performed using the LCAO approach. The Monkhorst-Pack scheme [19] with 16×16×16 *k*-point mesh for the BZ sampling and 32×32×32 *k*-point Gilat [25] net for the calculation of the Fermi energy and density matrix have been used here.

When modeling the UN(001) surface, we have used symmetric slabs consisting of five atomic layers with regularly alternating uranium and nitrogen atoms [15]. Plane wave computational formalism requires the use of an artificial slab translation in a vertical direction with a period called the *vacuum gap*. The magnitude of the latter (38.8 Å for five-layer UN slab), was found large enough to exclude the interaction between the repeated slabs for all single slab models studied using the PAW approach The slabs in the LCAO calculations have been really two-dimensional. The optimized lattice constant (4.87 for PAW *VASP vs.* 4.81 Å for LCAO *CRYSTAL* calculations) has been used in all further calculations, with an error within 2% of the experimental value (4.89 Å) [1]. Only ferromagnetic UN ground state has been considered in this study as the energetically most preferable state at low temperatures. The calculations of UN bulk structure suggest the magnetic moment on the U cation ~ 1 $\mu_B$. Thus, for five-layers slab the total magnetic moment of a 2×2 2D supercell (containing 20 U cations and 20 N anions) in both approaches has been fixed at 20 $\mu_B$.

To simulate the O atom adsorption, we have used the same supercell model with a periodic adsorbate distribution. These supercells with the 2×2 extension of surface translation vectors correspond to the atomic O coverage of 0.25 ML. To reduce computational efforts, we have considered symmetric two-sided arrangement of oxygen adatoms (Fig. 1). We have simulated two configurations of atomic adsorption: O atop the surface U cation or N anion (Fig. 1) with the complete structural optimization. For PAW calculations on the O/UN(001) interface using 3D slab model, we should also check whether the vacuum gap of 38.8 Å for a five-layer slab of uranium nitride [15] is large enough for the models additionally containing adsorbed O atoms from both sides.

The binding energy $E_{bind}$ of adsorbed oxygen $O_{ads}$ was calculated with respect to a free O atom:

$$E_{bind} = \frac{1}{2}\left(E_{tot}^{UN} + 2E_{tot}^{O_{triplet}} - E_{tot}^{O/UN}\right), \qquad (1)$$

where $E_{tot}^{O/UN}$ is the total energy of a fully relaxed O/UN(001) slab for $O_{ads}$ positions atop either the N or U surface ions, $E_{tot}^{O_{triplet}}$ and $E_{tot}^{UN}$ the energies of an isolated O atom in the ground (triplet) state and of a pure relaxed slab. In PAW calculations of free O atom, the cubic box with the same periodicity as for the O/UN(001) and UN(001) 3D slabs has been used. The factor ½ before brackets appears since the substrate is modeled by slab with the two equivalent surfaces and $O_{ads}$ is positioned symmetrically with respect to the surfaces.

Due to a mixed metallic-covalent nature of the chemical bonding in UN [10-14], we expect a high affinity of $O_{ads}$ towards the UN(001) substrate. The binding energy *per* O adatom is expected to be closer to that on a regular O/Al(111) and (001) metallic interfaces (~10 eV) [26] than on semiconducting O/SrTiO$_3$(001) interfaces (with two possible SrO- or TiO$_2$-terminations) (~2 eV) [27]. Indeed, we have obtained in the *VASP* calculations the binding energies of 6.9 and 5.0 eV *per* O adatom atop the surface U or N ions, respectively, accompanied with 0.9-1.2 *e* charge transfer from the surface towards the O adatom (Tables 1 and 2). The positively charged surface U cation goes outwards, to the adsorbed O atom whereas in the O configuration atop the N anion the latter is strongly displaced from the adsorbed O atom due to a mutual repulsion.

The corresponding results of *VASP* and *CRYSTAL* calculations based on the two very different methods demonstrate a good qualitative agreement for O adatom properties atop the surface U ion (Table 1) in all properties: the binding energies (3D slab models usually underestimate this parameter due to a weak repulsion between the adjacent polarized slabs), atomic displacements and even effective charges (which are calculated using the very different Mulliken (LCAO) and Bader (PAW) procedures).

An analysis of the difference electron charge redistributions for both configurations of $O_{ads}$ (Fig. 2) confirms that the O adatom forms a strong chemical bonding with the surface U cation which could be considered as one-site complex. In the case of O adatom atop the surface N anion this is rather multi-center adsorption complex involving four adjacent surface U ions. As follows from Table 1, these cations mostly contribute to the high O binding energy atop the N anion.

Adsorption of $O_{ads}$ atop the surface N or U ions on the UN(001) surface leads to appearance of the specific oxygen bands in the density of states (DOS) (Fig.3) as compared to DOS for a pure UN(001) surface [15]. For oxygen atop the surface U cation, O *2p* states overlap with the U *6d* and with a well-pronounced tail of U *5f* states in the region of the N *2p* valence band (-2 to -4 eV). This indicates once more a strong oxygen chemical bonding (chemisorption) on U typical for metal surfaces. However, when O is located atop N, the U *5f* contribution in this energy region diminishes whereas N *2p* states are considerably pushed down to smaller energies, due to N anion repulsion from negatively charged O adatom.

Summing up, the results obtained here for oxygen interaction with UN surfaces demonstrate strong chemisorption typical for metallic surfaces and could serve as the first important step in understanding the initial stage of the oxidation mechanism. The excellent agreement of the results obtained using two very different first principles methods supports their reliability. We continue the study of O$_2$ dissociation and the diffusion path of $O_{ads}$ on both perfect and defective UN(001) substrates, which is aimed at understanding atomistic mechanism of oxidation by means of substitution of surface N ions for O ions.

**Acknowledgements**

The authors kindly thank P. Van Uffelen, R. Caciuffo, D. Gryaznov, V. Kashcheyevs, A. Kuzmin, M. Losev, Yu. Mastrikov, S. Piskunov for fruitful discussions and for valuable help with the calculations. This study was partly supported by the Service Contract 205343-2006-07 F1ED KAR LV between ITU and ISSP, Riga and EC Framework 7 Project F-Bridge. D.B. gratefully acknowledges support from the European Social Fund (ESF).

Table 1. The calculated binding energy ($E_{bind}$), the distance between O and surface U cation ($d_{O-U}$), the effective atomic charges ($q$), and vertical ($\Delta z$) U and N displacements from the surface plane for adatom position atop the surface U (Fig. 1). The effective charges of U and N ions on the pure surface are equal to +1.63 $e$ for surface U cation and -1.55 $e$ for surface N anion in LCAO 5-layer slab calculations as well as +1.66 $e$ for surface U cation and -1.63 $e$ for surface N anion in PAW 5-layer slab calculations [15].

| Method of calculation | $E_{bind}$, eV | $q$(O), $e$ | $q$(U1), $e$ | $q$(U2), $e$ | $q$(U3), $e$ | $q$(N), $e$ | $d_{O-U}$, Å | $\Delta z$(U1), Å | $\Delta z$(U2), Å | $\Delta z$(U3), Å | $\Delta z$(N), Å |
|---|---|---|---|---|---|---|---|---|---|---|---|
| LCAO[a] | 8.3 | -0.89 | 1.97 | 1.66 | 1.62 | -1.56 | 1.87 | +0.15[b] | -0.07 | -0.11 | -0.04 |
| PAW[c] | 6.9 | -1.04 | 1.96 | 1.86 | 1.83 | -1.60 | 1.91 | +0.135[b] | -0.02 | -0.04 | -0.05 |

[a] LCAO-PBE calculations performed with *CRYSTAL-2006* code
[b] positive sign corresponds to atom displacement outward the substrate
[c] PAW-PW91 calculations performed with *VASP-4.6* code

Table 2. The calculated parameters for O atom adsorption atop the surface N anion[a]. See caption and footnotes of Table 1 for explanation.

| Method of calculation | $E_{bind}$, eV | $q$(O), $e$ | $q$(N1), $e$ | $q$(N2), $e$ | $q$(N3), $e$ | $q$(U), $e$ | $d_{O-N}$, Å | $\Delta z$(N1), Å | $\Delta z$(N2), Å | $\Delta z$(N3), Å | $\Delta z$(U), Å |
|---|---|---|---|---|---|---|---|---|---|---|---|
| PAW | 5.0 | -1.20 | -1.44 | -1.56 | -1.59 | -1.56 | 2.19 | -0.64 | +0.065 | +0.06 | +0.10 |

[a] atomic positions of U and N ions are reversed as compared to those shown in Fig. 1.

**Figure captions**

Fig. 1. A model of two-sided periodic adsorption of O atoms (0.25 ML) atop the surface U cations. Numbers enumerate non-equivalent interfacial atoms.

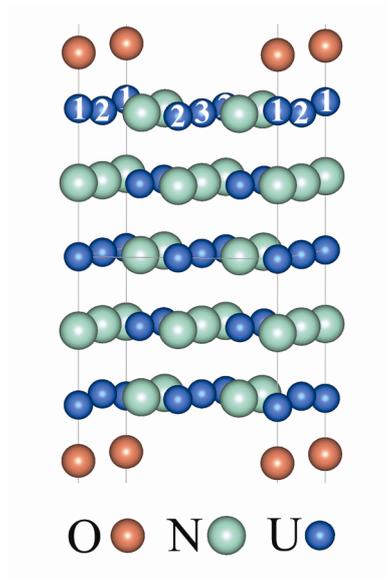

Fig. 2. The difference electron density maps $Dr(\mathbf{r})$ (the total density of the interface minus the densities of substrate and adsorbate with optimized interfacial geometry) for the O adatoms atop the surface: (a) N anions and (b) U cations on the UN(001) surface obtained using results of PAW calculations. Solid (red) and dashed (blue) isolines correspond to positive and negative electron density, respectively. Isodensity increment is 0.003 $e$ Å$^{-3}$.

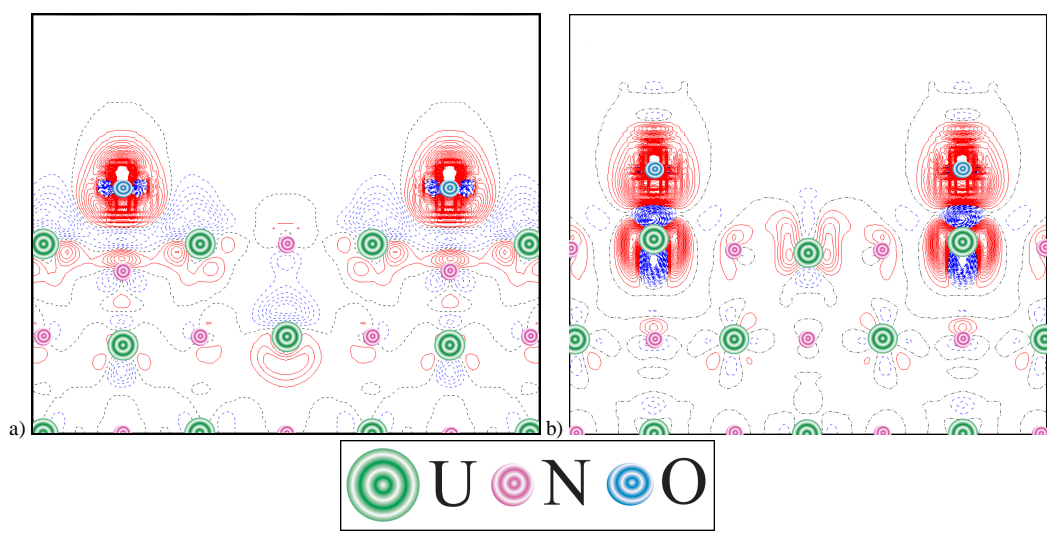

Fig. 3. The total and projected densities of states for O adsorption atop the N anion (a) and the U cation (b) obtained using results of PAW calculations. In the former, we consider the orbital projections of N anion under O atom and one of four nearest neighbouring U cations (Fig. 2a). Analogously, the lower plot presents the orbital projections of U cation beneath adatom and one of four nearest N anions. The largest peaks have been normalized to the same value, whereas a convolution of individual energy levels has been plotted using the Gaussian functions with a half-width 0.2 eV.

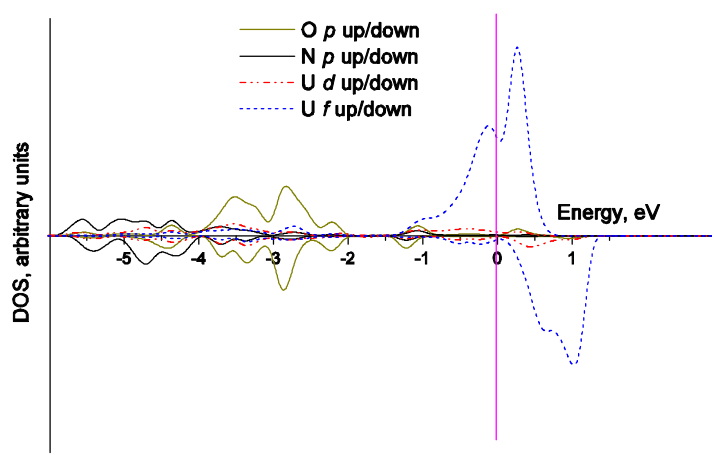

a)

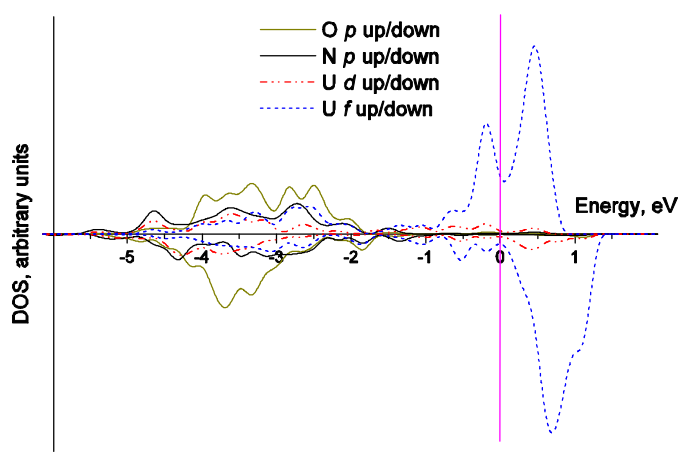

b)